\documentclass[10pt, twocolumn,
							 superscriptaddress,
							 english,
							 prb,
							 showpacs,
							 floatfix,
							 aps
							]{revtex4-2}
\usepackage{graphicx} 
\usepackage[utf8]{inputenc}
\usepackage{amsmath}
\usepackage{amssymb}
\usepackage{graphicx}
\usepackage{xspace}
\usepackage{xfrac}
\usepackage{xcolor}
\usepackage[backref=none,bookmarksnumbered=true,bookmarks=true,bookmarksopen=true,colorlinks=true,
citecolor=blue,linkcolor=blue,anchorcolor=green,urlcolor=blue,unicode=false]{hyperref}
\usepackage{ulem}

\newcommand{\twoH}{$2H$-NbSe$_2$}
\newcommand{\nbse}{NbSe$_2$}
\newcommand{\didv}{\ensuremath{\mathrm{d}I/\mathrm{d}V}\xspace}

\date{\today}

\begin{document}

\title{Direct signatures of $d$-level hybridization and dimerization in magnetic adatom chains on a superconductor}

\author{Lisa M. R\"{u}tten}
\affiliation{\mbox{Fachbereich Physik, Freie Universit\"at Berlin, 14195 Berlin, Germany}}

\author{Eva Liebhaber}
\affiliation{\mbox{Fachbereich Physik, Freie Universit\"at Berlin, 14195 Berlin, Germany}}

\author{Ga\"el Reecht}
\affiliation{\mbox{Fachbereich Physik, Freie Universit\"at Berlin, 14195 Berlin, Germany}}

\author{Kai Rossnagel}
\affiliation{\mbox{Institut für Experimentelle und Angewandte Physik, Christian-Albrechts-Universit\"at zu Kiel, 24098 Kiel, Germany}}
\affiliation{\mbox{Ruprecht Haensel Laboratory, Deutsches Elektronen-Synchrotron DESY, 22607 Hamburg, Germany}}

\author{Katharina J. Franke}
\affiliation{\mbox{Fachbereich Physik, Freie Universit\"at Berlin, 14195 Berlin, Germany}}

\begin{abstract}
      
Magnetic adatom chains on superconductors provide a platform to explore correlated spin states and emergent quantum phases. Using low-temperature scanning tunneling spectroscopy, we study the distance-dependent interaction between Fe atoms on \twoH. While single atoms exhibit four Yu-Shiba-Rusinov states and partially occupied $d$ levels consistent with a $S=2$ spin state, the spin is quenched when two Fe atoms reside in nearest neighbor lattice sites, where the $d$ levels of the atoms hybridize. The non-magnetic dimer configuration is stable in that dimerization persists in chains with weak interactions among the dimers. Thus, the spin-state quenching has important implications also for Fe chains. While even-numbered chains are stable and non-magnetic, odd-numbered chains host a single magnetic atom at one of the chain's ends, with its position being switchable by voltage pulses.
Our findings emphasize the role of interatomic coupling in shaping quantum ground states and suggest that engineering alternating hopping amplitudes analogous to the Su-Schrieffer-Heeger model may offer a pathway to realizing topological systems.

\end{abstract}

\maketitle
\section{Introduction}

Chains of magnetic atoms on superconductors constitute fascinating systems for the design of topological superconductivity and correlated one-dimensional states \cite{Yazdani2023, Choi2019}. Exchange interactions of the unpaired adatom spins with the underlying superconductor induce Yu-Shiba-Rusinov (YSR) states inside the superconducting gap of the substrate \cite{Yu1965, Shiba1968, Rusinov1968}. If two atoms are located sufficiently close to each other such that their YSR wave functions overlap, they may hybridize and form symmetric and antisymmetric linear combinations \cite{Rusinov1968, Flatte2000, Yao2014a, Hoffman2015, Ruby2018, Schmid2022}. In long chains these hybridized states evolve into bands which may host topological superconductivity if they are partially filled \cite{Pientka2013, Pientka2015}. For even closer spacing of adatoms hybridization and band formation may also happen directly in the atoms' $d$ levels. In case of an odd number of spinless bands crossing the Fermi level with either a non-collinear spin texture or strong spin-orbit coupling in the bulk, these systems may also exhibit topological superconductivity and Majorana end states \cite{NadjPerge2013, Klinovaja2013, Braunecker2013}. 

Indeed, signatures of Majorana zero modes were first observed in ferromagnetic chains of self-assembled Fe atoms on Pb(110) \cite{NadjPerge2014, Ruby2015chains, Pawlak2016} contrary to structurally similar Co chains \cite{Ruby2017}, highlighting the role of the number of $d$ derived bands crossing the Fermi level. Chains of densely spaced adatoms were also created by precise positioning of the atoms using the tip of a scanning tunneling microscope (STM), where the evolution of the YSR states was followed atom by atom but with little consideration of the $d$ levels \cite{Kim2018, Schneider2021, Friedrich2021, Mier2021, Schneider2022, Schneider2023, Wang2023}. Chains with larger interatomic spacings also showed signatures of YSR band formation where the $d$ level interactions are probably negligible \cite{Ding2021, Kuester2021a, Kuester2022, Liebhaber2022, Rutten2025a}. 

Common to all these experiments is an equidistant spacing of the atoms in the chains. Hence, the interaction and hopping amplitude between neighboring atoms can be assumed to be uniform. However, introducing a modulation in the hopping strength by deliberately varying the spacing or coupling between atoms would emulate the physics of the Su-Schrieffer-Heeger (SSH) model \cite{Su1979}. This model, originally developed to describe polyacetylene, predicts the emergence of topologically protected edge states in chains with alternating bond strengths. Realizing an SSH-like scenario with magnetic atoms on a superconducting substrate could be an additional approach to creating topological phases. 

The most natural way of realizing of alternating hopping strength in an atomic chain is by placing the atoms at alternating distances. However, the realization is constrained in that adatoms can typically be positioned only at integer multiples of the substrate’s lattice constant.

Here, we investigate the coupling strength between Fe atoms on \twoH\ by systematically varying the distance from nearest-neighbor sites to a few lattice constants using the tip of a scanning tunneling microscope. We observe an abrupt change in the interaction mechanism when the separation is reduced from two lattice constants to one. At the shortest distance, direct hybridization of the Fe $d$-orbitals becomes possible, whereas at larger separations, coupling occurs only through substrate-mediated interactions. Notably, this direct coupling leads to a complete quenching of the Fe dimer's magnetic moment.

As the chain length increases, we find that the atoms spontaneously dimerize, with even-numbered chains forming paired units, while odd-numbered chains exhibit an unpaired atom at one end that retains its magnetic moment and consequently exhibits YSR states.
Following the evolution of the $d$ levels from the single atom to the chain also suggests weak $d$ level coupling between dimers. While our chain thus self-tunes into dimerization with alternating hopping, the weak inter-dimer coupling is insufficient to support a Su–Schrieffer–Heeger (SSH)–like topological phase.

\section{Experimental methods}
We use a JT-scanning tunneling microscope from SPECS working at 1.1\,K for all experiments. We enhance our energy resolution beyond the Fermi-Dirac limit by using superconducting, Nb-coated W or NbTi tips. A Nb single crystal was cleaned by standard sputter and flash-anneal cycles until an oxygen reconstructed surface was obtained and differential conductance (\didv) spectra showed the full superconducting gap of bulk Nb. The tip was then repeatedly indented into the crystal until the \didv\ spectra showed approximately twice the superconducting gap width of Nb, revealing the presence of a superconducting gap at the tip. The same tip was then used on the \twoH\ crystals. We indicate tip gaps in \didv spectra by gray shaded areas and state their precise sizes in the figure captions. Smaller tip crashes were performed on both the Nb and the \nbse\ surface to sharpen and stabilize the tip.

We cleaved the \twoH\ crystals in ultra-high vacuum (UHV) using sticky tape and transferred them into the STM. We evaporated Fe atoms directly into the STM with the temperature remaining below 12\,K. We precisely positioned the atoms by laterally approaching the tip to an atom at a set point of a few nanoampere (1\,nA to 7\,nA depending on the tip apex) at 4\,mV and tracking its motion as jumps in the current and $z$ traces while dragging the atom across the surface. Upon reaching the desired position, we release the atom by returning to standard measuring set points. 

We record \didv spectra using two different protocols. For spectra across the superconducting gap we fix the tip at a specific height by switching off the feedback loop at the set point indicated in the figure captions. We then ramp the bias voltage while recording the \didv\ signal using a standard lock-in technique. For spectra at higher energies that do not span the superconducting gap, we keep the feedback loop activated during the measurement, such that the tip is retracted as the bias voltage increases. The \didv\ signal is obtained via a lock-in amplifier with a modulation signal added to the bias voltage at a given frequency. This procedure allows measurements over a large bias voltage range with large signals and without risking destruction of the tip or adatom structure. 

In addition to standard constant-current topographies, we also record constant-\didv\ topographies. In this case the feedback loop is set to regulate on a chosen \didv signal instead of the current. The resulting image closely resembles a constant contour of the density-of-states, which has proven particularly efficient for probing hybridization of states \cite{Reecht2017}.

\section{Results}
In this work, we aim to investigate the evolution of the Fe $d$-level resonances as two atoms are brought into proximity and as longer chains are formed. Since $d$-level hybridization is closely connected to modifications in the Yu-Shiba-Rusinov states, a full understanding of the electronic and magnetic interactions between adatoms requires examining both features simultaneously. We therefore begin by analyzing the YSR states of Fe dimers at varying separations. While the YSR states of dilute Fe dimers on \twoH\ have been extensively studied in previous works \cite{Liebhaber2022, Rutten2024, Rutten2025, Rutten2025a}, we revisit them here to provide a complete and unified perspective. 
Topographic images of the \twoH\ crystal show the atomically resolved terminating Se lattice and an additional periodic height modulation of approximately three lattice constants $a$ resulting from the incommensurate charge-density wave (CDW) that coexists with superconductivity in \twoH\ at low temperatures. The CDW has been shown to affect the energies of the YSR states \cite{Liebhaber2020}, requiring not only the determination of the adsorption site of the Fe atoms on the atomic lattice of the substrate, but also with respect to the CDW. Detailed information on the specific adsorption geometries of all structures discussed in the following can be found in Supplementary Information (SI), Section S1.

\subsection{YSR states of Fe monomers and dimers}
\begin{figure}\centering	\includegraphics[width=0.95\linewidth]{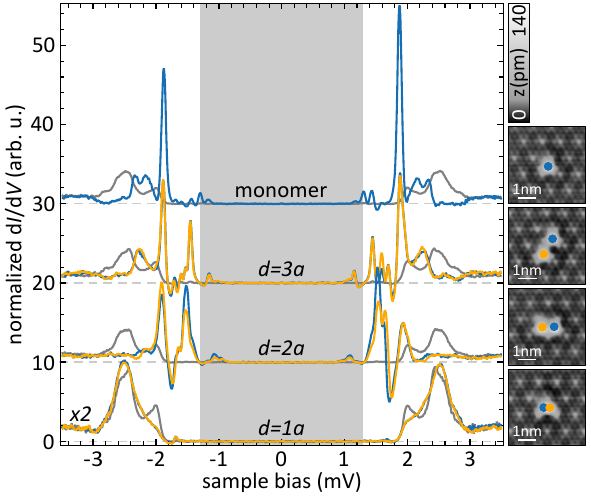}
	\caption{YSR states of Fe monomer and dimers at different spacing. The \didv spectra were recorded on the center of the Fe atoms as indicated by the colors on the corresponding topographic images shown on the right hand side. The interatomic spacings are given next to the spectra. Substrate spectra are shown in gray for comparison. The spectra recorded on the $1a$ dimer are multiplied by two for better visibility. 
	$\Delta_{\mathrm{tip}}\approx$~1.30\,mV (gray shaded area in spectra). Set points: topographies 10\,mV, 100\,pA; spectra: 5\,mV, 750\,pA; lock-in modulation V$_\mathrm{rms}$=15\,$\mu$V.
	}
	\label{fig:YSR}
\end{figure}

We begin with the Fe monomer, for which a differential conductance spectrum and atomic-resolution topographic image are shown in the top row of Fig.\,\ref{fig:YSR}. Consistent with previous studies, the monomer exhibits four YSR states that we ascribe to the crystal-field split $d$ levels with each of the singly-occupied levels inducing a YSR state \cite{Liebhaber2020}. Additional resonances at $\pm1.69$\,mV and $\pm1.59$\,mV originate from a nearby atom, as detailed in SI, Section S1.
When a second Fe atom (orange in Fig.\,\ref{fig:YSR}) is brought close to the original monomer (blue in Fig.\,\ref{fig:YSR}) at separations of only a few lattice constants ($d = 3a$ and $d = 2a$; second and third rows of Fig.\,\ref{fig:YSR}, respectively), the spectra display clear signatures of YSR hybridization. These include an increased number of resonances and shifts in their energies, consistent with earlier reports \cite{Liebhaber2022, Rutten2024}. In the $d = 2a$ configuration, the peaks appear broader and with shoulders, which we interpret as several split YSR states overlapping in energy \cite{Rutten2025}.

Interestingly, once the atoms are brought into nearest-neighbor sites, i.e., at a distance of $a$ that we refer to as a ``dense" dimer (bottom row of Fig.\,\ref{fig:YSR}), the individual atoms are no longer resolved in the topographic images. At the same time, all YSR states associated with the dimer atoms disappear from the differential conductance spectra. (The small remaining resonances within the superconducting gap originate from a different nearby atom, as discussed in SI, Section S1.)

The abrupt disappearance of YSR states upon forming a dense dimer may arise from several mechanisms. Most notably, $d$ level hybridization drastically changes the energy levels and their occupation, which may lead to the formation of a singlet state $S=0$. The absence of any magnetic moment consequently suppresses the YSR states. Similar singlet formation has been observed for closely spaced Mn dimers on Cu$_2$N and MoS$_2$ \cite{Hirjibehedin2006, Trishin2023}. This scenario of direct coupling of the $d$-derived states will be examined in more detail below.

However, beyond this straightforward interpretation, other factors must be carefully considered, in particular changes in the exchange coupling strength arising from variations in adsorption site or local strain. For instance, the atoms may attract each other and become slightly displaced from their original adsorption sites. This scenario is possible as the lattice constant of Fe crystals is smaller than the spacing between the hollow sites of \nbse\ that are occupied by the atoms. Strain-induced modifications of YSR states have been reported in previous experiments \cite{Wang2023}, and density functional theory suggests that for dense Cr dimers on $\beta$-Bi\textsubscript{2}Pd, a combination of atomic displacement and antiferromagnetic spin alignment is likely \cite{Choi2018}. If the Fe atoms in the dense dimer are indeed lifted from their original monomer adsorption sites, the exchange coupling to the substrate would likely be reduced. 

\subsection{$d$ levels of Fe monomer and dimers}
\begin{figure*}\centering	\includegraphics[width=0.95\linewidth]{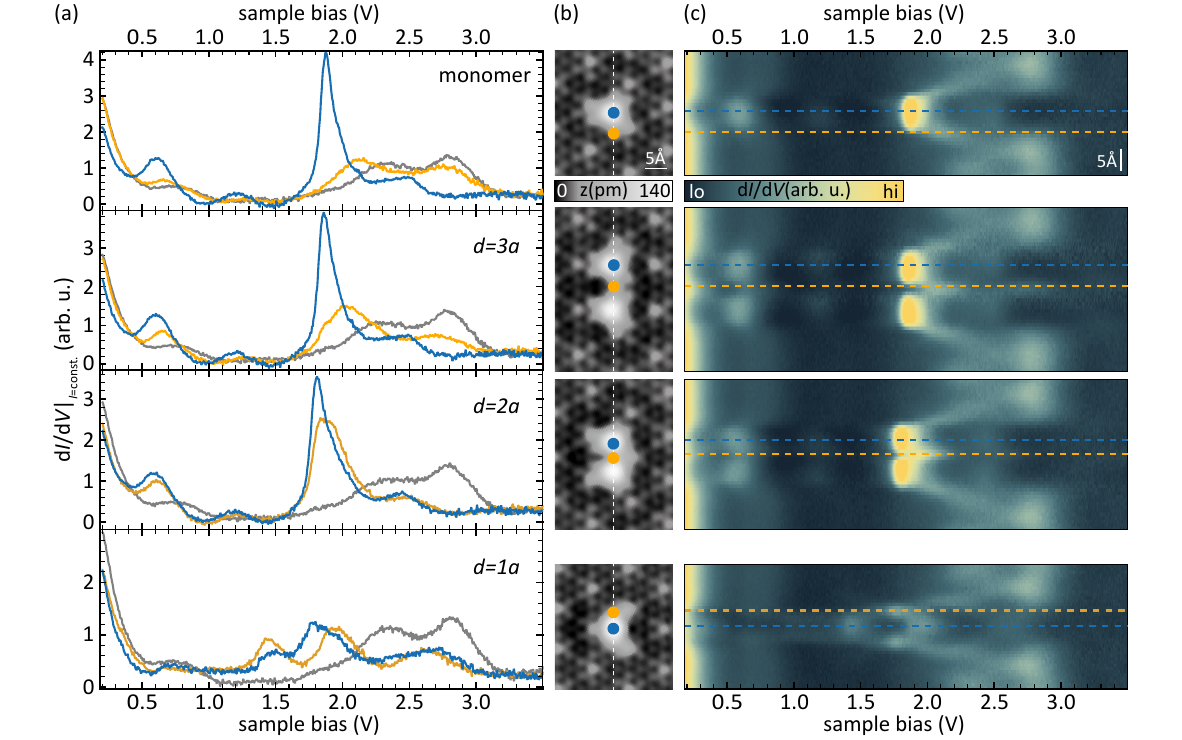}
	\caption{Spectral signature of $d$ levels. (a) Constant-current \didv spectra of the monomer and three differently spaced dimers (spacings indicated in each panels are valid for each row). (b) STM topographies of the corresponding adatom structures where the positions at which the spectra in (a) were taken are indicated by color coded dots. (c) False-color plots of constant-current spectra recorded along the dashed lines in the corresponding panel in (b). The vertical axes of topographies and color plots are aligned. The spectra plotted in (a) are indicated by dashed lines of corresponding color.
		Set points: (a), (c) 200\,pA; (b) 10\,mV, 100\,pA; V$_\mathrm{rms}$=5\,mV. 
	}
	\label{fig:dlevels}
\end{figure*}

To possibly distinguish these scenarios, we investigate the change of the $d$ levels in the different dimers. To do so, we recorded constant-current \didv spectra as detailed in the methods section. As YSR states originate from unpaired spins in partially occupied $d$ levels, the corresponding $d$ states should show up as resonances at both positive and negative bias voltages. Here, we focus on the positive bias range, where the lower tunneling barrier allows for better resolution of individual resonances compared to the occupied states at negative bias. For spectra at negative bias see SI, Section S3.
Figure\,\ref{fig:dlevels}a shows selected spectra of the Fe monomer and dimers at various adatom spacings. For comparison, spectra recorded on the bare surface are shown in gray. The positions at which the spectra were taken are marked by color-coded dots in the topographic images in Fig.\,\ref{fig:dlevels}b. Spectra of the monomer and the dimers at $2a$ and $3a$ appear very similar. We observe three distinct resonances at approximately 0.60\,V, 1.20\,V, and 1.85\,V with the latter exhibiting a shoulder on its high-energy side. These four features are consistent with the presence of four partially filled $d$ levels, as expected from the four observed YSR states. At a distance of $2a$, the most intense resonance shifts downward in energy by about 50\,meV compared to the monomer, indicating some weak interactions among the atoms. With only minor changes in the spectra, we classify this and all larger spacings as dilute limit.

Spectra recorded on the $1a$ dimer (bottom row of Fig.\,\ref{fig:dlevels}) differ drastically from those on the monomer and the $2a$ and $3a$ dimers. Unlike the distinct resonances seen in the dilute cases, spectra on the center of the $1a$ dimer are nearly flat up to 1.3\,V with the first clear resonance at approximately 1.45\,V, another one at 1.95\,V and a broader feature at 2.5\,V (orange spectrum). Hence, four resonances in the unoccupied states of the single atom are replaced by three unoccupied ones on the dimer. The edge of the dimer is distinct from the center with a resonance at approximately 1.80\,V (blue spectrum). The pronounced changes in the $d$-level spectra for the $1a$ dimer, as opposed to the relatively minor changes in the dilute $2a$ and $3a$ dimers, suggest a strong interaction between the $d$ levels. Given the close spacing, we refer to this case as the dense limit.

To elucidate the nature of this interaction, we examine the spatial distribution of the constant-current \didv spectra along the dimer axis. The constant-current spectra along the white dashed lines in the corresponding topographic images (Fig.\,\ref{fig:dlevels}b) are shown as color plots in Fig.\,\ref{fig:dlevels}c. 
For the Fe monomer and atoms in the dilute dimers the three resonances mentioned above are localized at the center of the atoms. 
Slightly off the atoms, the most intense resonances seems drastically reduced in intensity and dispersing toward higher energy, finally merging with the bulk bands at 2.3\,eV. In between the atoms of the $2a$ and $3a$ dimer, these states overlap and lead to a broad feature with no space to further disperse.

As already inferred from the individual \didv spectra, the $d$ levels are drastically different for the $1a$ dimer. The resonance at 1.45\,V is centered between the atoms of the dimer, consistent with a bonding orbital. In contrast, the resonance at 1.80\,V is localized at the outer edges of the dimer with no intensity in between. This distribution resembles an anti-bonding orbital. The resonances at 1.95\,V and 2.5\,V are also localized between the atoms, potentially representing additional bonding orbitals. Furthermore, we observe states dispersing toward higher energies at the dimer edges. The similarity of these features across all dimers suggests a substrate-related origin, with the energy shifts likely influenced by the local potential of the adatoms. 

\begin{figure}\centering	\includegraphics[width=0.95\linewidth]{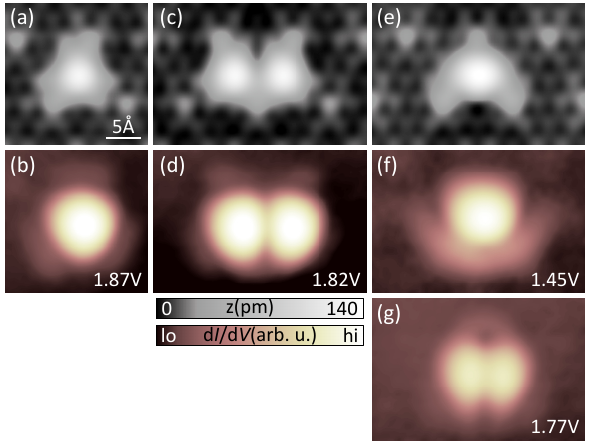}
	\caption{Mapping the hybridization of $d$ levels. (a), (c), (e) Constant-current STM topographies of an Fe monomer, a $2a$ dimer and a $1a$ dimer, respectively. (b), (d), (f), (g) Constant-\didv\ STM topographies recorded of the same areas depicted above the respective panel at the energies indicated in the bottom right corner.		
		Set points: (a), (c), (e) 10\,mV, 100\,pA; (b), (d), (f), (g): bias voltage given in each panel, lock-in voltage=1\,V; (b), (d), (g): V$_\mathrm{rms}$=30\,mV, (f): V$_\mathrm{rms}$=50\,mV.
	}
	\label{fig:DoSmaps}
\end{figure}

Bonding and anti-bonding orbitals originate from the symmetric and anti-symmetric linear combination of the crystal-field split $d$ levels. This picture can be validated by analyzing constant-\didv\ topographies, which are most appropriate for bringing out the absence/presence of nodal planes at the corresponding bias voltages. We start by discussing such a constant-\didv\ image of the Fe monomer. Figure\,\ref{fig:DoSmaps}a and b show a constant-current topography at 10\,mV and a constant-\didv\ topography at 1.87\,V of the same area, respectively.  In the constant-current image (Fig.\,\ref{fig:DoSmaps}a), the atom appears with a triangular shape, likely resulting from hybridization between the Fe $d$ states and the surrounding Se atoms forming the triangular adsorption site \cite{Trishin2021}. 
The corresponding constant-\didv\ image in Fig.\,\ref{fig:DoSmaps}b displays large intensity within a rounded triangle shaped region centrally around the atom. Note that this triangle is mirrored relative to the one seen in the constant-current image. In addition to this central shape there are elevated regions parallel to the sides of the central triangle, reflecting the spatially extended orbital from hybridization with the Se atoms.

Figure\,\ref{fig:DoSmaps}c and d show an analogous data set recorded on the $2a$ dimer. The atoms are sufficiently close-spaced for parts of the monomer features to overlap as already visible from the constant-current topography (Fig.\,\ref{fig:DoSmaps}c). In the constant-\didv\ topography at 1.82\,V (Fig.\,\ref{fig:DoSmaps}d) the main features that we observed on the monomer (i.e., a rounded triangular shape of the high intensity surrounded by smaller features that are approximately parallel to the triangle's sides) appear deformed but recognizable on the $2a$ dimer. The almost preserved shape together with the minimal energy shift compared to the monomer, corroborate the absence of strong $d$ level interactions. 

In case of the $1a$ dimer the individual atoms can no longer be distinguished in the constant-current topography (Fig.\,\ref{fig:DoSmaps}e). This pronounced change in the topographic signal is accompanied by significant alterations in the spectral features compared to both the monomer and the dilute dimers. We focus here on constant-\didv\ images of the previously identified symmetric state at 1.45\,V (Fig.\,\ref{fig:DoSmaps}f) and the antisymmetric state at 1.77\,V (Fig.\,\ref{fig:DoSmaps}g).
 The image of the state at 1.45\,V reveals a single bright trapezoidal feature centered on the dimer, likely resulting from the spatial compression of features observed on the $2a$ dimer. Below this bright region, a lower-intensity, bowl-shaped feature is visible, which may originate from the overlapping side-parallel features associated with each atom. However, the upper side-parallel features seen in the monomer and $2a$ dimer are not clearly discernible here. Crucially, no nodal line is observed between the atoms, supporting the interpretation of this state as a symmetric linear combination of monomer orbitals.

In contrast, Fig.\,\ref{fig:DoSmaps}g, recorded at 1.77\,V over the same area, clearly shows a nodal line centered on the dimer, indicating an antisymmetric combination of the monomer states. Additionally, the two bright lobes separated by the nodal line are more narrow on their bottom ends than on their top ends which may be reminiscent of the triangular shapes. The two bright lobes on either side of the nodal line are narrower at the bottom, potentially reminiscent of the triangular features observed on the monomer. A faint feature above the lobes may relate to the top-side-parallel feature seen on the monomer.

Among the states we analyzed, the one at 1.77\,V is the only one that clearly exhibits antisymmetric character. (Additional constant-\didv\ images are provided in SI, Section S4.) Other antisymmetric states may lie beyond our energy window or be obscured by overlapping symmetric states. It is important to emphasize that hybridization significantly alters the energy level alignment and state occupation. Due to the limited resolution at negative bias voltage, we cannot completely track the hybridization of all crystal-field split levels.
 
However, from drastic changes of the $d$ level spectra, we conclude that the absence of YSR states in the dense dimer is most likely due to hybridization of the adatoms' $d$ levels and their occupation leading to a spin singlet such that only full or empty but no partially filled levels are present in the $1a$ dimer. 

\subsection{Dense Fe chains}
The hybridization of $d$ derived states provides a promising basis for band formation. To observe related signatures, we extend the $1a$ dimer by adding additional atoms to form dense Fe\textsubscript{$n$} chains, where $n$ indicates the number of atoms in the chain. Here, we show representative data on an even (Fe\textsubscript{6}) and odd numbered (Fe\textsubscript{7}) chain.

\begin{figure*}\centering	\includegraphics[width=0.95\linewidth]{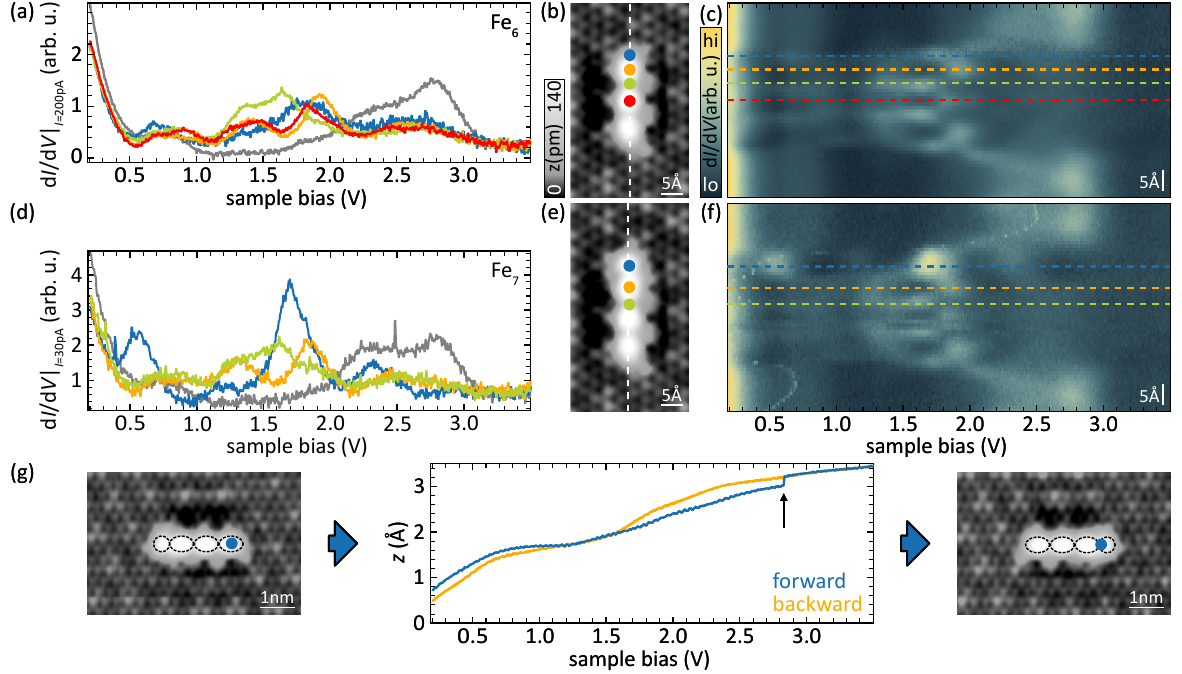}
	\caption{Dense Fe chains. (a), (d) Constant-current spectra at different positions along the Fe\textsubscript{6} and Fe\textsubscript{7} chain, respectively. (b), (e) Constant-current topographic images with the positions at which the spectra in (a) were recorded marked by color coded dots. (c), (f) Color plots of Ccnstant-current spectra recorded along the white dashed lines in (b), (e) with the spectra shown in (a), (d) indicated by horizontal dashed lines of corresponding color. The spectra in (d), (f) exhibit higher noise levels because they were recorded at a lower current set point. (g) Switching of the location of the single end of the Fe\textsubscript{7} chain. Topographic images show the chain before and after switching the location of the single end by measuring the constant-current spectrum shown in the center at the position indicated by the blue dot. The central panel shows the $z$ trace recorded during a constant-current spectrum, starting from 200\,mV. The switching event is highlighted by the black arrow.
		Set points: (a), (c): 200\,pA; (b), (e), topographies in (g): 10\,mV, 100\,pA; (d), (f): 30\,pA; constant-current spectrum in (g): 100\,pA; V$_\mathrm{rms}$=5\,mV.
	}
	\label{fig:dlevels67}
\end{figure*}

The topography of the Fe\textsubscript{6} chain exhibits three bright lobes, each of them resembling the shape of the $1a$ dimer (Fig.\,\ref{fig:dlevels67}b). The topographic appearance, thus, suggests dimerization along the chain. To resolve the consequences in the electronic structure, we monitor the $d$ levels in the same energy range as for the dimer above. Individual spectra are shown in Fig.\,\ref{fig:dlevels67}a with their corresponding position marked by the color-coded dots on the topography. The spectra at the terminating dimer (both on the center and edge) closely resemble those on an isolated dimer. Only the spectrum between two dimers (green) shows a slightly different intensity distribution between 1.2\,V and 1.8\,V as compared to the edge spectrum (blue). Similarly, the resonance, that is found at approximately 1.9\,V centrally on the terminating dimer (orange) is found at lower energy on the middle dimer of the Fe\textsubscript{6} chain (red).
 
The difference between the chain's edges and its bulk is also visible in the color plot of spectra recorded along the chain (Fig.\,\ref{fig:dlevels67}c). Generally states of the dimer within the bulk of the chain appear at lower energies compared to the dimers at the chain edges. The electronic structure of a dimer, thus, seems to be preserved with only weak interaction among the dimers. Consequently, dimerization of the Fe atoms seems to inhibit the formation of extended $d$ derived bands.

To probe the robustness of dimerization, we attach another atom to the chain, thus creating an odd numbered Fe\textsubscript{7} chain. A topographic image of this chain is shown in Fig.\,\ref{fig:dlevels67}e. The three dimers already observed for the Fe\textsubscript{6} chain are still visible, while there is an additional single protrusion at the upper end of the chain (around the blue dot). To investigate the electronic structure upon addition of the Fe atom, we record constant-current \didv\ spectra on the single-atom end as well as on the dimer (color coded dots are shown in Fig.\,\ref{fig:dlevels67}d). The blue trace in Fig.\,\ref{fig:dlevels67}d was recorded on the single end of the Fe\textsubscript{7} chain. Notably, the spectral shape resembles the one observed on an isolated atom albeit slightly downshifted in energy, indicating a small but non-negligible interaction.
The green and the orange spectra were recorded on the first dimer next to the single atom and at equivalent positions as on the dimer in Fig.\,\ref{fig:dlevels67}a. While their overall spectral shape resembles those on a single $1a$ dimer but downshifted by $\approx$0.1\,V, this dimer now behaves similar to the dimer in the center of the Fe\textsubscript{6} chain. 

This conclusion is corroborated by a color plot of densely spaced constant-current spectra recorded centrally along the complete chain (Fig.\,\ref{fig:dlevels67}f, with the location marked by a white dashed line in Fig.\,\ref{fig:dlevels67}e). The intense resonance of the single end is clearly visible in this plot and causes the absence of mirror symmetry around the center of the chain. At the same time the upper dimer of the chain (orange traces) exhibits the same features as the middle dimer indicating that it behaves as part of the chain's bulk by just adding a single atom to be its neighbor. 

It is worth noting that the spectra of the Fe\textsubscript{7} chain were recorded using a lower current set point compared to the other structures. This reduction of the current set point was necessary due to the bistable nature of Fe\textsubscript{odd} chains, where the position of the single end atom can switch between both ends of the chain. This process is illustrated in Fig.\,\ref{fig:dlevels67}g. The left panel shows a constant-current topography with the single atom located at the left hand side of the chain. Upon recording a constant-current \didv\ spectrum at the position indicated by the blue dot, the single-atom end can be transferred to the right side of the chain. This switching event is captured in the $z$ trace shown in the central panel, where a sudden jump in tip height at approximately 2.8\,V (black arrow) indicates the transition, observed at a current set point of 100\,pA. After the measurement, the end atom appears on the right side of the chain, as seen in the topographic image in the right panel.

\begin{figure}\centering	\includegraphics[width=0.95\linewidth]{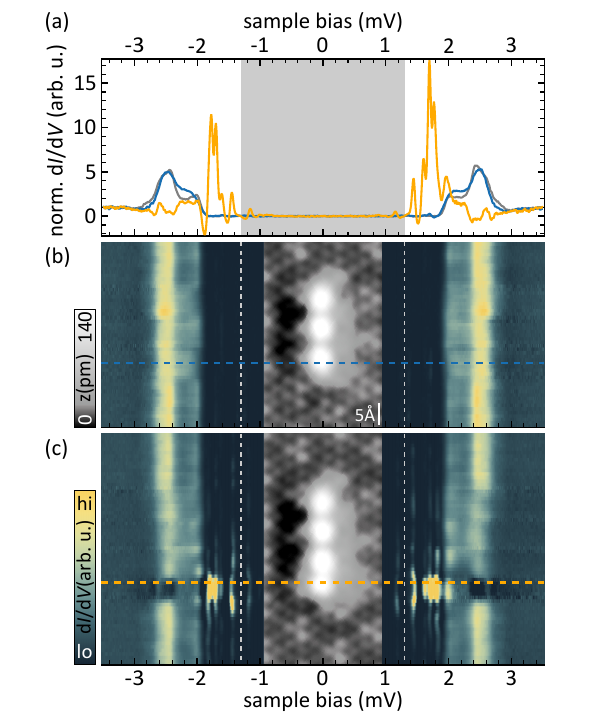}
	\caption{Low energy spectra of an Fe\textsubscript{6} and an Fe\textsubscript{7} chain. (a) Selected spectra recorded on a dimerized end of an Fe\textsubscript{6} chain and on the single-atom end of an Fe\textsubscript{7} chain with a spectrum of the substrate shown in gray. (b), (c) Color plots of spectra recorded centrally along the Fe\textsubscript{6} and the Fe\textsubscript{7} chain, respectively. The insets show topographic images of the respective chain that are aligned with the positions where the individual spectra were recorded. Dashed, color coded, horizontal lines mark the spectra shown in (a).
	$\Delta_{\mathrm{tip}}\approx$~1.30\,mV (gray shaded area in spectra); set points: topographies 10\,mV, 100\,pA; spectra: 5\,mV, 750\,pA, V$_\mathrm{rms}$=15\,$\mu$V. 
	}
	\label{fig:YSR67}
\end{figure}

To probe the magnetic properties of the odd-numbered chain, we examine the superconducting gap region for YSR states of the Fe\textsubscript{6} and an Fe\textsubscript{7} chain for YSR states. Figure\,\ref{fig:YSR67}a shows \didv\ spectra recorded on a dimerized end of the Fe\textsubscript{6} chain (blue) and on the single-atom end of the Fe\textsubscript{7} chain. The spectrum recorded on the Fe\textsubscript{6} chain closely resembles that of a dense dimer and does not exhibit any YSR states. Contrary, the spectrum of the Fe\textsubscript{7} chain's single end shows four YSR states well inside the superconducting gap. While this again indicates four unpaired electrons similar to the isolated Fe atom, their energies are very different. This shift is in agreement with the energetic differences also observed between the $d$ levels of an isolated monomer and the single end of the Fe\textsubscript{7} chain.

Figure\,\ref{fig:YSR67}b and c show color plots of spectra recorded centrally along the Fe\textsubscript{6} and Fe\textsubscript{7} chain, respectively. Here, the vertical dashed white lines indicate the superconducting gap of the tip and the horizontal dashed lines highlight the spectra of corresponding color shown in Fig.\,\ref{fig:YSR67}a. While there are no YSR states visible all along the Fe\textsubscript{6} chain, the YSR states of the Fe\textsubscript{7} chain originate from its single end and can be resolved all along the chain due to the long-range nature of the YSR wave functions.


\section{Conclusions}

In conclusion, we have systematically investigated the distance-dependent interaction between Fe atoms on a superconducting \twoH\ substrate. As YSR states are long-ranged on this quasi-two-dimensional material, hybridization of these states is possible at distances of a few atomic lattice sites of the substrate. In this limit the $d$ levels are largely unperturbed. This is in strong contrast to Fe atoms located in nearest-neighbor sites, where the $d$ levels hybridize and lead to the formation of a spin singlet. The YSR states are abruptly suppressed at this distance. Upon increasing chain length, we find that the dimer structure remains as a stable unit independent of chain length with weak interaction among neighboring dimers. The stability of the dimer units leads to a pronounced difference between odd- and even-numbered chains. While the even-numbered chains exhibit no magnetic moment, the isolated edge atom in the odd-numbered chains remains with its high spin. We further demonstrated that Fe\textsubscript{odd} chains are bistable regarding the position of their single end. We found  that the position of the single end can readily be changed using voltage pulses from the STM tip underscoring the critical role of the energetic balance within the chain. While our system features robust dimers with non-magnetic ground state, our findings highlight the potential of designing magnetic dimer structures to explore correlated quantum states and spin-dependent phenomena. Furthermore, engineering non-uniform hopping amplitudes may possibly offer new routes towards realizing a Su-Schrieffer-Heeger model with topologically nontrivial edge states.

\acknowledgements

We gratefully acknowledge financial support by the Deutsche Forschungsgemeinschaft (DFG, German Research Foundation) through Projects No. 277101999 (CRC 183, Project No. C03) and No. 328545488 (CRC 227, Project No. B05). LMR acknowledges membership in the International Max Planck Research School ``Elementary Processes in Physical Chemistry".


%

\clearpage

\newcommand{\beginsupplement}{%
	\setcounter{table}{0}
	\renewcommand{\thetable}{S\arabic{table}}%
	\setcounter{figure}{0}
	\renewcommand{\thefigure}{S\arabic{figure}}%
	\setcounter{equation}{0}
	\renewcommand{\theequation}{S\arabic{equation}}%
	\setcounter{section}{0}
	\renewcommand{\thesection}{S\arabic{section}}%
}

\setcounter{figure}{0}
\setcounter{section}{0}
\setcounter{equation}{0}
\setcounter{table}{0}

\onecolumngrid

\newcommand{\vsigma}{\mbox{\boldmath $\sigma$}}

\beginsupplement
\section*{{Supplementary Information}}

\maketitle 

\section{Positions of the dimer atoms}
\begin{figure}[h]
\centering	\includegraphics[width=\linewidth]{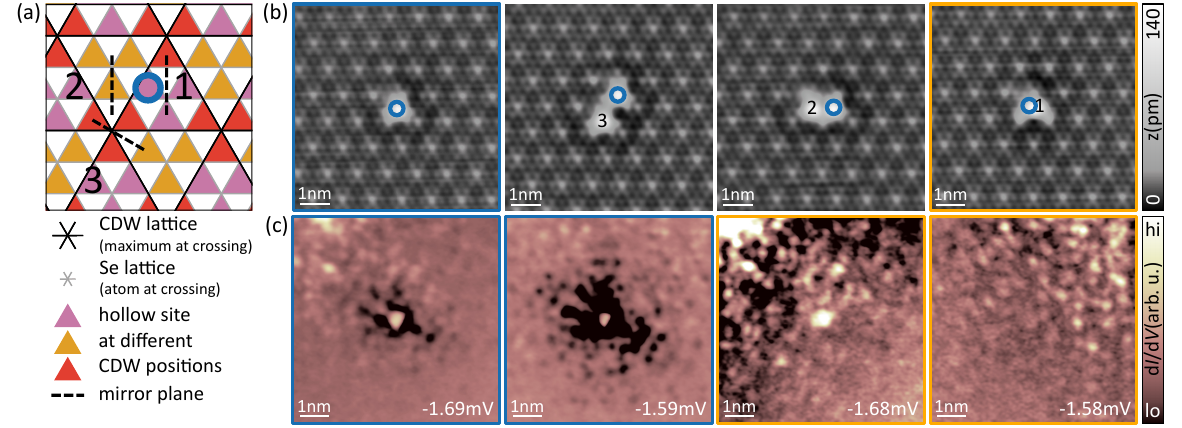}
	\caption{Determination of adsorption sites. (a) Schematic adsorption geometry and (b) larger scale topographic images of the monomer and all dimers discussed in the main text. (c) Constant-contour \didv\ maps of the monomer (blue frames) and the $1a$ dimer (orange frames) at the energies at which we observes weak in-gap resonances on the $1a$ dimer.
    $\Delta_{\mathrm{tip}}\approx$~1.30\,mV. Set points: topographies 10\,mV, 100\,pA; \didv\ maps: 5\,mV, 750\,pA; lock-in modulation V$_\mathrm{rms}$=15\,$\mu$V.
    }
	\label{fig:S1}
\end{figure}

The Yu-Shiba-Rusinov states of individual Fe atoms adsorbed in hollow sites of \nbse\ without a Nb atom underneath are strongly affected by the charge-density wave (CDW) that coexists with superconductivity in \twoH\ \cite{SLiebhaber2020}. All dimers considered in this work (see e.g. Fig.\,1 and Fig.\,2 of the main text) reside on a so-called chalcogen-centered CDW background, where CDW maxima coincide with Se atoms of the surface. This scenario is depicted in the sketch shown in Fig.\,\ref{fig:S1}a. Here, the Se lattice of the topmost layer is visualized by gray lines, where a Se atom is located at each intersection. A similar lattice for the $\sim 3 \times 3$ CDW is overlayed in black, where crossings represent CDW maxima. Using these two lattices we can distinguish different atomic hollow sites with respect to the CDW. We color code the different hollow sites without a Nb atom underneath according to their position with respect to the CDW as indicated by colored triangles. The resulting schematic allows to directly identify equivalent sites and the symmetries present for any combination of sites. 

We chose the adsorption sites for the dimer such that all separations $1a$, $2a$ and $3a$ exhibit mirror symmetry perpendicular to the dimer axis. Starting with an Fe atom located on a site marked by the blue circle on a purple triangle, the second atom is then placed at purple sites labeled 1, 2, 3 for the $1a$, $2a$ and $3a$ dimer, respectively. 
The correct placements can be verified by overlaying the CDW and Se lattices to the topographic images shown in Fig.\,\ref{fig:S1}b, where we use the same labeling of the atoms as introduced. 

As we mentioned in the main text, there is a faint peak within the superconducting energy gap in the spectrum measured on the $1a$ dimer (Fig. 1, lowest panel in the main text). To show that this is not a feature of the dimer but rather belongs to another atom, we closely inspect the constant-contour \didv\ maps in this region. 
Figure\,\ref{fig:S1}c depicts constant-contour \didv\ maps of the monomer (blue frames) and the $1a$ dimer (orange frames) at the energies of those small YSR resonances. These maps clearly show the long-ranged intensity scattering into the scan frame from the upper left rather than scattering patterns originating from the adatom structure under investigation. Thus, we can neglect these features in our analysis of the YSR states.

\section{Absence of spin excitations}

\begin{figure}\centering	\includegraphics[width=\linewidth]{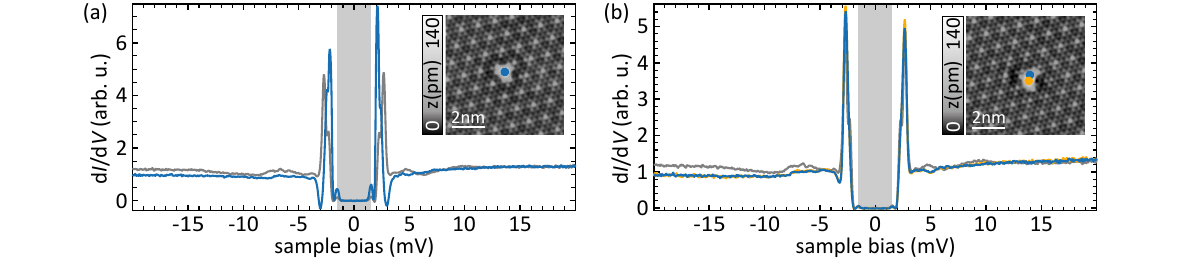}
	\caption{Low-energy spectra. (a) Spectra recorded on the monomer and (b) the $1a$ dimer at the positions indicated in the inset topographies. Spectra of the bare substrate are shown in gray for comparison. 
	 $\Delta_{\mathrm{tip}}\approx$~1.52\,mV; set points: 5\,mV, 750\,pA, lock-in modulation V$_\mathrm{rms}$=15\,$\mu$V. 
     }
	\label{fig:S2}
\end{figure}

If the spins of two atoms in a dimer form a spin singlet one may observe the singlet-triplet excitation in \didv\ spectra as the opening of a new inelastic tunneling channel at a threshold energy \cite{SHirjibehedin2006, STrishin2023}.
Figure\,\ref{fig:S2} shows spectra of the monomer (a) and the $1a$ dimer (b) over an energy range, in which such an excitation should be found if accessible in our system. However, neither the monomer nor the $1a$ dimer show any features other than those of the substrate outside the coherence peaks. We may speculate that the lifetime of such an excitation is too small on the superconductor to show a pronounced feature.

\section{Constant-current spectra at negative bias voltages}
\begin{figure}\centering	\includegraphics[width=\linewidth]{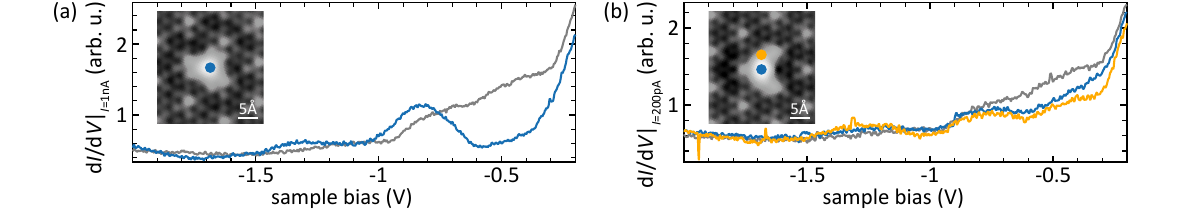}
	\caption{Constant-current spectra at negative bias voltage. (a) Spectra of the monomer and (b) the $1a$ dimer recorded at the positions indicated by color coded dots in the inset topographies. Spectra of the bare substrate are shown in gray for comparison. 
	Set points: (a) $I=1$\,nA; (b) $I=200$\,pA; lock-in modulation V$_\mathrm{rms}$=5\,mV; insets 10\,mV, 100\,pA. 
	}
	\label{fig:S3}
\end{figure}

Figure\,\ref{fig:S3} shows constant-current spectra of the monomer (a) and the $1a$ dimer (b) recorded at negative bias voltages, where (partially) occupied $d$ levels may be probed. For the monomer we observe the most pronounced resonance around $-0.85$\,V and an additional small feature around $-1.35$\,V. Furthermore, there is a resonance around $-0.40$\,V, that was not observed on the monomer. The spectra recorded on the $1a$ dimer, thus, differ significantly from that recorded on the monomer especially above $-1$\,V. This is in agreement with the hybridization picture laid out in the main manuscript with a rearrangement of the $d$ levels and different occupation. Unfortunately, the background spectra also exhibit some features in the given energy range, greatly hampering an interpretation of this data.

\section{Additional constant-\didv images}
\begin{figure}\centering	\includegraphics[width=\linewidth]{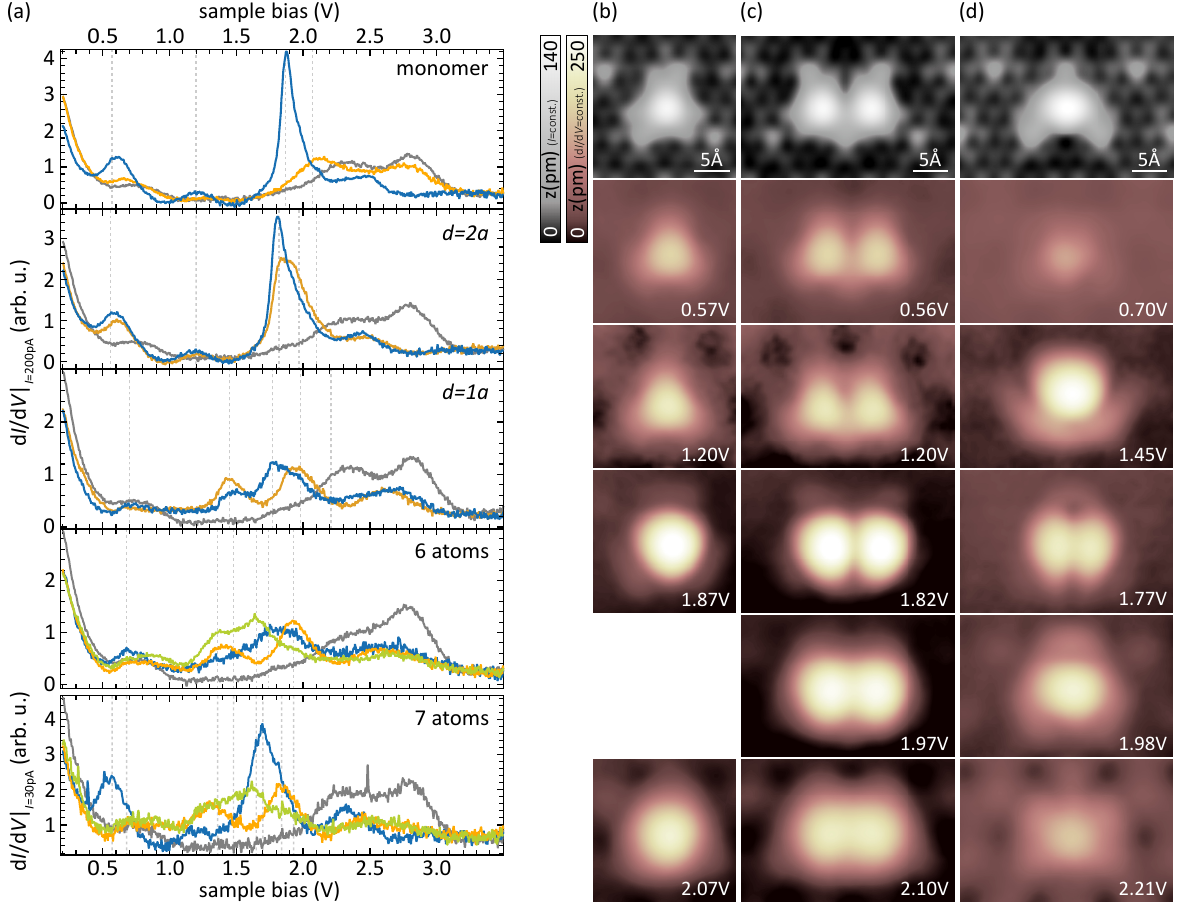}
	\caption{Additional data on monomer and dimers. (a) Constant-current spectra of all adatom structures reproduced from the main text. Vertical dashed lines indicate the energies, at which constant-\didv images shown in (b)-(d) were recorded. Constant-current (top row) and constant-\didv (rest) images of (b) the monomer, (c) the $2a$ dimer and (d) the $1a$ dimer.
	Set points: (a) Given in axis label; (b)-(d) top row: 10\,mV, 100\,pA, rest: bias voltage given in each panel, lock-in voltage=1\,V, panels at 1.20\,V and 1.45\,V: V$_\mathrm{rms}$=50\,mV, rest: V$_\mathrm{rms}$=30\,mV.
     }
	\label{fig:S4}
\end{figure}

\begin{figure}\centering	\includegraphics[width=\linewidth]{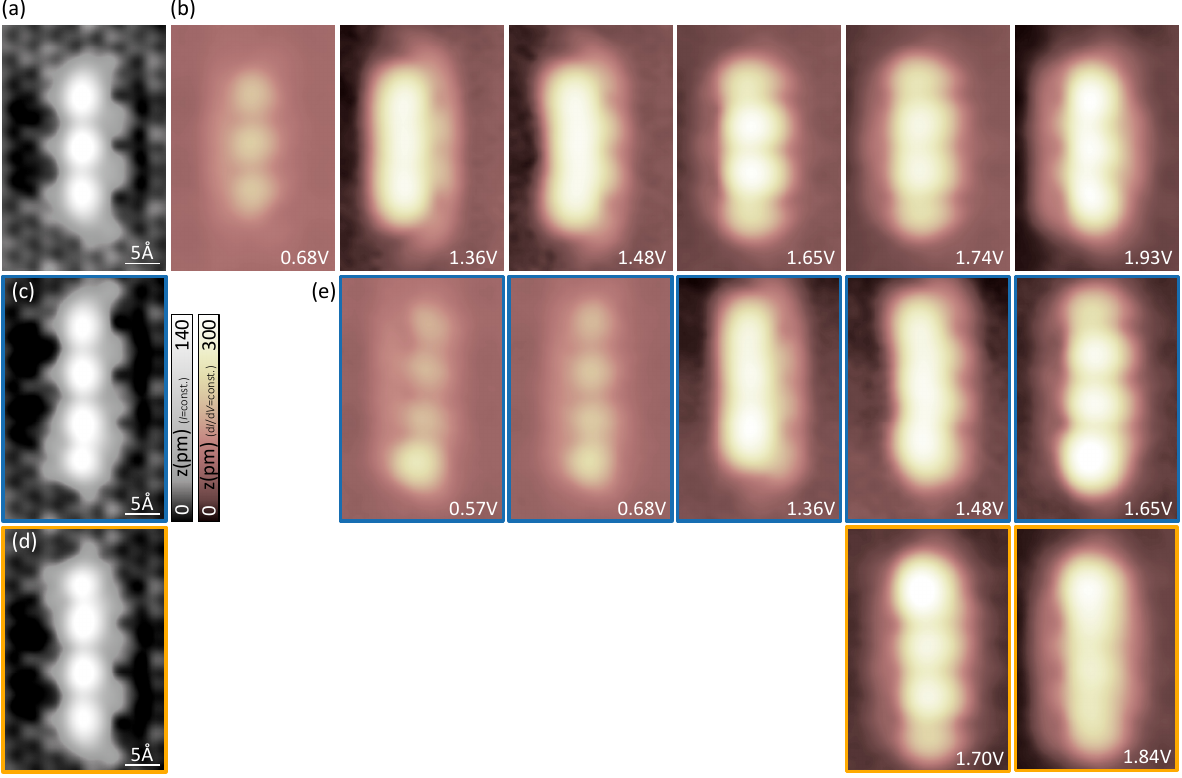}
	\caption{Additional data on Fe chains. (a), (c), (d) Constant-current and (b), (e) constant-\didv topographies of the Fe\textsubscript{6} (a), (b) and the Fe\textsubscript{7} chain (c)-(e). For the Fe\textsubscript{7} chain the position of the single end changed between some of the images from the single end being at the bottom in (c) and the top row of (e) (blue frames), to the single end being at the top in (d) and the bottom row of (e) (orange frames).
	Set points: (a), (c), (d) 10\,mV, 100\,pA; (b), (e) bias voltage given in each panel, lock-in voltage=1\,V, V$_\mathrm{rms}$=30\,mV.
	}
	\label{fig:S5}
\end{figure}

Figures\,\ref{fig:S4} and \ref{fig:S5} show additional constant-\didv images of all structures discussed in the main text. Additionally, constant-current spectra from the main text are reproduced in Fig.\,\ref{fig:S4}a, where vertical dashed lines indicate the energies at which the constant-\didv images were recorded.

\def\urlprefix{}
  \def\url#1.{}

\end{document}